\documentclass[amsmath,amssymb,nofootinbib]{revtex4}
\usepackage{amsmath,amssymb,graphics,epsfig,subfigure}
\usepackage{color}
\newtheorem{theorem}{Theorem}

\begin{document}
\title{Partition of kinetic energy and magnetic moment in dissipative diamagnetism}
\author{Jasleen Kaur\footnote{jk14@iitbbs.ac.in}, Aritra Ghosh\footnote{ag34@iitbbs.ac.in}, and Malay Bandyopadhyay\footnote{malay@iitbbs.ac.in}}
\affiliation{School of Basic Sciences, Indian Institute of Technology Bhubaneswar,\\   Jatni, Khurda, Odisha, 752050, India}
\date{\today}

\begin{abstract}
In this paper, we analyze dissipative diamagnetism, arising due to dissipative cyclotron motion in two dimensions, in the light of the quantum counterpart of energy equipartition theorem. We consider a charged quantum particle moving in a harmonic well, in the presence of a uniform magnetic field, and coupled to a quantum heat bath which is taken to be composed of an infinite number of independent quantum oscillators. The quantum counterpart of energy equipartition theorem tells us that it is possible to express the mean kinetic energy of the dissipative oscillator as a two-fold average, where, the first averaging is performed over the Gibbs canonical state of the heat bath while the second one is governed by a probability distribution function $P_k(\omega)$. We analyze this result further, and also demonstrate its consistency in the weak-coupling limit. Following this, we compute the equilibrium magnetic moment of the system, and reveal an interesting connection with the quantum counterpart of energy equipartition theorem. The expressions for kinetic energy and magnetic moment are reformulated in the context of superstatistics, i.e. the superposition of two statistics. A comparative study of the present results with those obtained from the more traditional Gibbs approach is performed and a perfect agreement is obtained.
\end{abstract}
\maketitle

\section{Introduction}
In recent times, we have observed rapid advancements in the field of statistical mechanics
of dissipative quantum systems \cite{1}. In such kind of studies, the system of interest is usually coupled to a heat bath whose degrees of freedom are integrated out, leading to quantum dissipation. Typically, studies in the context of quantum thermodynamic properties of open quantum systems involve two
different approaches: (i) the first one is the usual Gibbs approach that focuses
on the partition function which can be computed from the weighted average of the values of the relevant observables at all possible phase points that lie on a constant time slice \cite{2a,2b}, and (ii) the other one is the Einstein approach, which is based on a quantum Langevin equation for the
subsystem \cite{3a,ford1988}. A nice aspect about this second method is that in addition to equilibrium quantities, non-equilibrium and approach-to-equilibrium properties can also be calculated \cite{singh,sdg1,hanggi,PRE79,sdg2,15,18,20}.\\

Based on the above-mentioned two approaches, there have been various studies exploring the quantum thermodynamic behavior of small systems
in the presence of finite quantum dissipation \cite{15,a2,hanggi,a4}. It has been observed that the presence of
quantum dissipation is actually helpful in restoring third law of thermodynamics for the archetype cases
of a damped harmonic oscillator and a free quantum Brownian particle \cite{a2}. In addition, the effect of quantum dissipation on the much studied problem of Landau diamagnetism \cite{malay,malay1,malay2}, and the
analysis of thermodynamic properties and thermodynamic laws can be found in \cite{18,20}.  Further, the effect of dissipation, and several subtle issues on the low- and high-temperature behaviors of the specific heat in the context of dissipative diamagnetism have been reported in \cite{PRE79,sdg2}. Recently, there have been several studies on the quantum analogue of energy equipartition theorem \cite{jarzy1,jarzy2,jarzy3,jarzy4,superstat,jarzy5,jarzy6,kaur,kaur1,kaurfermi}. It has been shown
that unlike the classical case, the average energy of an open
quantum system can be interpreted as being the sum of contributions from individual bath degrees of freedom distributed over the entire frequency spectrum. Such contributions are governed by certain probability distribution functions which are very sensitive to the parameters of the dissipative quantum system \cite{jarzy2,kaur}.\\

In the present study, we focus on making a bridge between the study of thermodynamic behavior of nanoscale quantum systems and the recently invented quantum analogue of energy equipartition theorem for such systems. For this purpose, we consider an exemplary model of
dissipative diamagnetism: a charged quantum particle moving in a harmonic well, in the presence of a uniform external magnetic field, and coupled to a quantum heat bath which is taken to be composed of an infinite number of independent quantum oscillators. The magnetic response of a charged quantum particle has wide and important
relevance in Landau diamagnetism \cite{landau1,landau2,peirls} (see also \cite{malay}), the quantum Hall effect \cite{hall1,hall2}, atomic physics \cite{abcd1}, and two-dimensional electronic systems \cite{abcd2,abcd3,abcd4}, to name just a few. By considering
such a physically relevant model system, we ask whether the dissipative magnetic moment of the system can be interpreted in the light of the quantum energy equipartition theorem. If yes, is there a connection between the distribution function(s) which govern the mean magnetic moment and the mean kinetic energy (or the mean potential energy) of the system of interest? Further, a recent advancement in statistical physics is the concept of superstatistics i.e. the statistics of the
statistics \cite{SS1,SS2} (see also \cite{superstat}). Following this, in the present work, we also analyze dissipative diamagnetism from the point of view of superstatistics. Finally, we compare the results obtained from the quantum Langevin equation with the more traditional Gibbs approach and obtain a perfect agreement.\\

With this preamble, we present the organization of the paper as follows. In the next section, we introduce our model system and discuss the basic ingredients of the quantum Langevin equation. We then present two constituent theorems which shall be proved in later sections. In section (\ref{mainresultsection}), we compute the mean kinetic energy of the charged dissipative oscillator from the quantum Langevin equation, thereby proving theorem-(\ref{theorem1}) stated in section (\ref{modelsec}). Subsequently, in subsection (\ref{partitionsubsec}), we revisit the quantum counterpart of energy equipartition theorem and discuss some important features of the kinetic energy distribution. In section (\ref{magneticmomentsec}), we prove theorem-(\ref{theorem2}) and discuss its physical significance in the context of dissipative diamagnetism. A connection between the quantum counterpart of energy equipartition theorem and dissipative diamagnetism is pointed out. section (\ref{superstatsection}) is devoted to the analysis of our results from the point of view of superstatistics (see for example, \cite{superstat}). Thereafter, in section (\ref{seriessection}), we express the equilibrium magnetic moment of the oscillator in the form of an infinite series and discuss the role of the system parameters on the behavior of the magnetic moment. The equivalence between the Einstein method and the Gibbs approach is established. We conclude our paper in section (\ref{conclusions}).

\section{Model, method, and observables}\label{modelsec}
In this section we briefly describe the model of interest and the corresponding quantum Langevin equation method. Let us consider a quantum charged particle of mass \(m\) and charge \(e\) confined to a harmonic potential with frequency \(\omega_0\), and acted upon by a transverse uniform magnetic field \(B\). Further, it is linearly coupled to a heat bath which comprises of an infinite number of independent harmonic oscillators. Thus, the total Hamiltonian reads
\begin{eqnarray}\label{H}
   H &=& \frac{(\mathbf{p} - \frac{e}{c} \mathbf{A})^2}{2m} + \frac{m \omega_0^2 \mathbf{r}^2}{2}  + \sum_{j=1}^N\bigg[\frac{\mathbf{p}_j^2}{2m_j} + \frac{1}{2}m_j \omega_j^2 \bigg( \mathbf{q}_j - \frac{c_j}{m_j \omega_j^2}\mathbf{r} \bigg)^2 \bigg],
\end{eqnarray}
where $\mathbf{p}$ and $\mathbf{r}$ are the momentum and position operators of the system, $\mathbf{p}_j$ and $\mathbf{q}_j$ are the corresponding variables for the \(j\)th  oscillator of the thermostat, and $\mathbf{A}$ is the vector potential. The usual commutation relations between coordinates and momenta hold. Integrating out the reservoir variables from the Heisenberg equations of motion and assuming that the system and the bath were in a product form of Gibbs canonical state initially \cite{kaur1,bez,canizaro}, one obtains a quantum Langevin equation (see for example \cite{ford1988} and references therein):
\begin{equation}\label{eqnm}
  m \ddot{\mathbf{r}}(t) + \int_{-\infty}^{t} \mu(t - t') \dot{\mathbf{r}}(t') dt' + m \omega_0^2 \mathbf{r}(t)-\frac{e}{c}(\dot{\mathbf{r}}(t) \times \mathbf{B}) = \mathbf{f}(t),
\end{equation}
where \(\mu(t)\) is the dissipation kernel, given by
\begin{equation}\label{mudef}
  \mu(t) = \sum_{j=1}^{N} \frac{c_j^2}{m_j \omega_j^2} \cos (\omega_j t) \Theta (t),
\end{equation}
 and \(\mathbf{f}(t)\) is an operator valued random noise whose spectral properties are characterized by the following symmetric correlation and the commutator:
\begin{eqnarray}
\langle \lbrace f_{\alpha}(t), f_{\beta}(t^{'}) \rbrace \rangle &=& \frac{2 \delta_{\alpha\beta}}{\pi}\int_{0}^{\infty}d\omega \hbar \omega {\rm Re}[\tilde{\mu}(\omega)] \coth\Big(\frac{\hbar\omega}{2k_BT}\Big) \cos \lbrack \omega(t-t^{'})\rbrack,  \label{symmetricnoisecorrelation} \\
\langle \lbrack f_{\alpha}(t), f_{\beta}(t^{'}) \rbrack\rangle &=& \frac{2\delta_{\alpha\beta}}{i\pi}\int_{0}^{\infty}d\omega \hbar \omega {\rm Re}[\tilde{\mu}(\omega)] \sin\lbrack \omega(t-t^{'})\rbrack. \label{noisecommutator} 
\end{eqnarray} In the above equations \(\tilde{\mu}(\omega)\) represents the Fourier transform of the friction kernel \(\mu(t)\). Here $\alpha$ and $\beta$ are being used to indicate Cartesian indices \(x\) and \(y\). The angular brackets in Eqs. (\ref{symmetricnoisecorrelation}) and (\ref{noisecommutator}) imply thermal averaging over the heat bath. Let us recall that the bath spectral function \(J(\omega)\) characterizing the spectral distribution of the bath degrees of freedom is defined as
\begin{equation}\label{Jdef}
  J(\omega)=\frac{\pi}{2}\sum_{j=1}^{N} \frac{c_j^2}{m_j \omega_j} \delta(\omega-\omega_j).
\end{equation}

\subsection{Observables}
In this subsection, we discuss the main quantities of interest in detail and convey our plan with the formulation of two constituent theorems. With the help of the above-mentioned model system, we revisit dissipative Landau diamagnetism exhibited by the above setup, and ask whether there is any connection between dissipative diamagnetism and the quantum counterpart of energy equipartition theorem studied in \cite{jarzy1,jarzy2,jarzy3,jarzy4,superstat,jarzy5,jarzy6,kaur,kaur1,kaurfermi}. As we shall subsequently show, the answer is affirmative. We begin our analysis by considering the mean kinetic energy of the dissipative magnetic system. The following result shall be proved:

\begin{theorem}\label{theorem1}
The mean kinetic energy of a two-dimensional charged oscillator of mass \(m\), electric charge \(e\), and placed in magnetic field \(\mathbf{B} = B \hat{z}\) in a state of thermal equilibrium with the quantum heat bath at temperature \(T\) can be expressed as
\begin{equation}\label{eqn1theorem1}
E_k =  \frac{1}{2\pi}\int_{-\infty}^{\infty}d\omega \mathcal{E}_k(\omega) \omega^2 [\Phi(\omega)+\Phi(-\omega)],
\end{equation} where \(\mathcal{E}_k(\omega) = \frac{\hbar \omega}{2} \coth (\frac{ \hbar \omega}{2k_B T})\) is the mean kinetic energy of a two-dimensional bath oscillator of frequency \(\omega\), and the function \(\Phi(\omega)\) is given by
\begin{eqnarray}\label{eqn2theorem1}
\Phi(\omega)&=& \frac{{\rm Re}[\tilde{\gamma}(\omega)]}{\Big[\Big(\omega^2-\omega_0^2-\omega\omega_c + \omega {\rm Im}[\tilde{\gamma}(\omega)]\Big)^2+(\omega {\rm Re}[\tilde{\gamma}(\omega)] )^2\Big]}. 
\end{eqnarray} Here, \(\tilde{\gamma}(\omega)\) is the Fourier transform of the friction kernel (per unit mass) appearing in the quantum Langevin equation, \(\omega_0\) is the system's eigenfrequency, and \(\omega_c = eB/mc\) is the cyclotron frequency.
\end{theorem}

Notice that \(\mathcal{E}_k(\omega)\) looks like the mean kinetic energy of a two-dimensional quantum oscillator, weakly coupled to a heat bath at temperature \(T\). The function \(\Phi(\omega)\) has been called the relaxation function in \cite{ford}. A similar statement as above can be proved for the potential energy, but for brevity, we do not pursue it in detail. In section (\ref{magneticmomentsec}), we shall prove the following result:

\begin{theorem}\label{theorem2}
The equilibrium magnetic moment \(M_z\) of a dissipative charged oscillator in two dimensions in a state of thermal equilibrium with the quantum heat bath at temperature \(T\) can be expressed as
\begin{equation}\label{magneticpartition}
M_z =  \frac{1}{2\pi }  \int_{-\infty}^{\infty}d\omega m(\omega)\omega^2 [\Phi(\omega)-\Phi(-\omega)],
\end{equation} where \(m(\omega) = -\mu_B \coth\Big(\frac{\hbar\omega}{2 k_B T}\Big)\) (not to be confused with mass) is the thermal Bohr magneton.
\end{theorem}
Thus, the equilibrium magnetic moment of the charged dissipative oscillator can be expressed as an integral over the bath spectrum. Its physical interpretation and a novel connection with the quantum counterpart of energy equipartition theorem shall be revealed in section (\ref{magneticmomentsec}). \\

Theorems (\ref{theorem1}) and (\ref{theorem2}) basically enumerate superstatistics in the frequency domain. This implies the superposition of two statistics \cite{SS1,SS2} i.e. they involve two-fold averaging: the first one is over the Gibbs canonical state for the thermostat oscillators while the second averaging is over randomly distributed frequencies $\omega$ of the thermostat oscillators according to suitable probability distribution functions.
For kinetic energy, the relation $\zeta =\mathcal{E}_{k}(\omega)= \frac{\hbar\omega}{2}\coth\big(\frac{\hbar\omega}{2k_BT}\big)$ enables us to define a new random variable $\zeta$ with a certain distribution function \(f_{k}(\zeta,T)\) in the energy representation \cite{superstat}. This will help us to interpret theorem-(\ref{theorem1}) in a new way. A similar analysis can be carried out for the magnetic moment [theorem-(\ref{theorem2})] whereby, Eq. (\ref{magneticpartition}) can be reformulated in the thermal Bohr magneton representation with some distribution \(f_m (m,T)\) (see subsection (\ref{ssmmsec})). 

\section{Partition of kinetic energy}\label{mainresultsection}
In this section, we shall compute the mean kinetic energy of the dissipative oscillator from the quantum Langevin equation [Eq. (\ref{eqnm})]. Although the quantum counterpart of energy equipartition theorem for a charged oscillator in a magnetic field was studied earlier in \cite{kaur}, we re-derive the expression for mean kinetic energy directly from the solution of the equation of motion (quantum Langevin equation) rather than invoking the fluctuation-dissipation theorem as has been implemented in \cite{jarzy5,kaur}. This shall help us to set-up our notation and subsequently, in the next section [section (\ref{magneticmomentsec})], we'll discuss its connection with the equilibrium magnetic moment. 

\subsection{Mean kinetic energy}
 For our convenience, let us define the variable \(Z = x + iy\). Then the solution to Eq. (\ref{eqnm}) can be expressed as (see also \cite{singh,sdg1})
\begin{equation}\label{Z}
 Z(t) = N \int_{0}^{t} d\tau \bigg[ e^{\omega_+ (t - \tau)} - e^{\omega_- (t - \tau)} \bigg] f (\tau),
\end{equation} where \(f(t) = f_x(t) + i f_y(t)\), and
\begin{equation}
	\omega_{\pm} = - \frac{[{\rm Re}[\tilde{\gamma}(\omega)] + i {\rm Im}[\tilde{\gamma}(\omega)] + i \omega_c]}{2} \pm \frac{1}{2} \sqrt{[{\rm Re}[\tilde{\gamma}(\omega)] + i {\rm Im}[\tilde{\gamma}(\omega) + i \omega_c]]^2- 4 \omega_0^2}, \hspace{5mm} N = \frac{1}{m (\omega_+ - \omega_-)}.
\end{equation} Here \(\tilde{\gamma}(\omega) = \tilde{\mu}(\omega)/m\). We now compute the kinetic energy of the system as\footnote{Here, `c.c' denotes complex conjugate.}
\begin{equation}
E_k(t) = \frac{m}{2} \langle \dot{x}(t)^2 + \dot{y}(t)^2 \rangle = \frac{m}{4} \langle \dot{Z}(t) \dot{Z}^\dagger (t) + {\rm c.c.} \rangle,
\end{equation} where from Eq. (\ref{Z}), we have
\begin{equation}\label{dotZ}
 \dot{Z}(t) = N \int_{0}^{t} d\tau \bigg[ \omega_+ e^{\omega_+ (t - \tau)} - \omega_- e^{\omega_- (t - \tau)} \bigg] f (\tau).
\end{equation}
 Thus, one can express the kinetic energy at time instant \(t\) in the following form:
\begin{eqnarray}
E_k(t) = \frac{m |N|^2}{4 \pi} \int_{-\infty}^\infty d\omega \hbar \omega \coth \bigg( \frac{ \hbar \omega}{2 k_B T} \bigg) && \Bigg[ \int_{0}^{t} d\tau \bigg( \omega_+ e^{\omega_+ (t - \tau) - i \omega \tau} - \omega_- e^{\omega_- (t - \tau) - i \omega \tau}\bigg) \\
&& \times \bigg( \omega^*_+ e^{\omega^*_+ (t - \tau) - i \omega \tau} - \omega^*_- e^{\omega^*_- (t - \tau) - i \omega \tau}\bigg) \Bigg],  \nonumber
\end{eqnarray}
or, in the steady state,
\begin{equation}\label{kineticenergysteadystate}
E_k =  \frac{1}{2 \pi} \int_{-\infty}^\infty d\omega \omega^2 \mathcal{E}_k (\omega) [\Phi(\omega) + \Phi(-\omega)],
\end{equation}
where \(\mathcal{E}_k (\omega) = \frac{\hbar \omega}{2} \coth (\frac{ \hbar \omega}{2 k_B T}) \) is the thermally-averaged kinetic energy of a two-dimensional bath oscillator of frequency \(\omega\), and
\begin{eqnarray}
\Phi(\omega) = m |N|^2  \bigg|\frac{1}{\omega - i \omega_-} - \frac{1}{\omega - i \omega_+}\bigg|^2 = \frac{{\rm Re}[\tilde{\gamma}(\omega)]}{\Big[\Big(\omega^2-\omega_0^2-\omega\omega_c + \omega {\rm Im}[\tilde{\gamma}(\omega)]\Big)^2+(\omega {\rm Re}[\tilde{\gamma}(\omega)] )^2\Big]}.  \label{Phi2}
\end{eqnarray}
Eq. (\ref{kineticenergysteadystate}) is identical to Eq. (\ref{eqn1theorem1}) and completes the proof of theorem-(\ref{theorem1}).

\subsection{Energy partition}\label{partitionsubsec}
We shall now discuss energy partition, or equivalently, the quantum counterpart of energy equipartition for the kinetic energy of the dissipative oscillator \cite{jarzy1,jarzy2,jarzy3,jarzy4,superstat,jarzy5,jarzy6,kaur}. The integral appearing in Eq. (\ref{kineticenergysteadystate}) can be rewritten as
\begin{equation}\label{kineticenergysteadystate1}
E_k =  \frac{1}{\pi} \int_{0}^\infty d\omega \omega^2 \mathcal{E}_k (\omega) [\Phi(\omega) + \Phi(-\omega)].
\end{equation}
The following result can be proven directly (see subsection (\ref{KEPE})):
\begin{equation}\label{omegaPhinormalize}
\int_{-\infty}^\infty \omega^2 \Phi(\omega) d\omega =  \pi,
\end{equation} which implies that
\begin{equation}\label{mnbvcxz}
\frac{1}{\pi} \int_{-\infty}^\infty \omega^2 \Phi(\omega) d\omega =  \frac{1}{\pi} \int_{0}^\infty \omega^2 [\Phi(\omega) + \Phi(-\omega)] d\omega = 1.
\end{equation}
Moreover, since \({\rm Re}[\tilde{\mu}(\omega)] > 0\), a requirement emerging from the second law of thermodynamics \cite{ford1988}, we may define
\begin{equation}\label{pkgenexp}
P_k(\omega) = \frac{1}{\pi} \omega^2 [\Phi(\omega) + \Phi(-\omega)],
\end{equation} which is essentially a probability distribution function, i.e. it is both positive definite and normalized on the interval \(\omega \in [0, \infty)\). Thus, we can express the mean kinetic energy of the dissipative oscillator as
\begin{equation}\label{Ekpart}
E_k = \int_0^\infty  \mathcal{E}_k (\omega) P_k(\omega) d\omega,
\end{equation} which is the quantum counterpart of energy equipartition theorem observed earlier \cite{jarzy1,jarzy2,jarzy3,jarzy4,superstat,jarzy5,jarzy6,kaur}. \\

So far our analysis has been kept general, suited for an arbitrary dissipation mechanism described by some choice of \(\tilde{\mu}(\omega)\) or \(\tilde{\gamma}(\omega)\), satisfying the requirements presented in \cite{ford1988}. However, for concreteness and simplicity, throughout the rest of this section, we opt for Ohmic dissipation for which \(\tilde{\gamma}(\omega) = \gamma_0\). Thus, we have
\begin{equation}\label{pkohmexp}
P_k(\omega) = \pi^{-1} \omega^2 \gamma_0 \Bigg[\frac{1}{\big[(\omega^2-\omega_0^2-\omega\omega_c)^2+(\omega \gamma_0 )^2\big]} + \frac{1}{\big[(\omega^2-\omega_0^2+\omega\omega_c)^2+(\omega \gamma_0 )^2\big]} \Bigg].
\end{equation} The probability distribution function has been plotted in figures-(1)-(4). Inspecting figures-(1) and (2), it becomes clear that although for small magnetic fields, the probability distribution function starts with a single peak, as the magnetic field gets larger, two distinct peaks emerge. Since from Eq. (\ref{Ekpart}), we can interpret \(\mathcal{E}_k(\omega) P_k(\omega) d\omega\) as the contribution to the mean kinetic energy of the dissipative oscillator, coming from bath degrees of freedom lying in the frequency interval between \(\omega\) and \(\omega + d\omega\), we conclude that there are up to two most probable frequencies such that bath degrees of freedom lying around those frequencies make the most significant contribution to the mean kinetic energy of the system. The magnetic field can be regarded as an external control parameter which has significant control over the location of the peaks, as well as over their height. Furthermore, in the absence of magnetic field, the two peaks coalesce into one. \\

\begin{figure}
	\centering
		\includegraphics[width=4.3in]{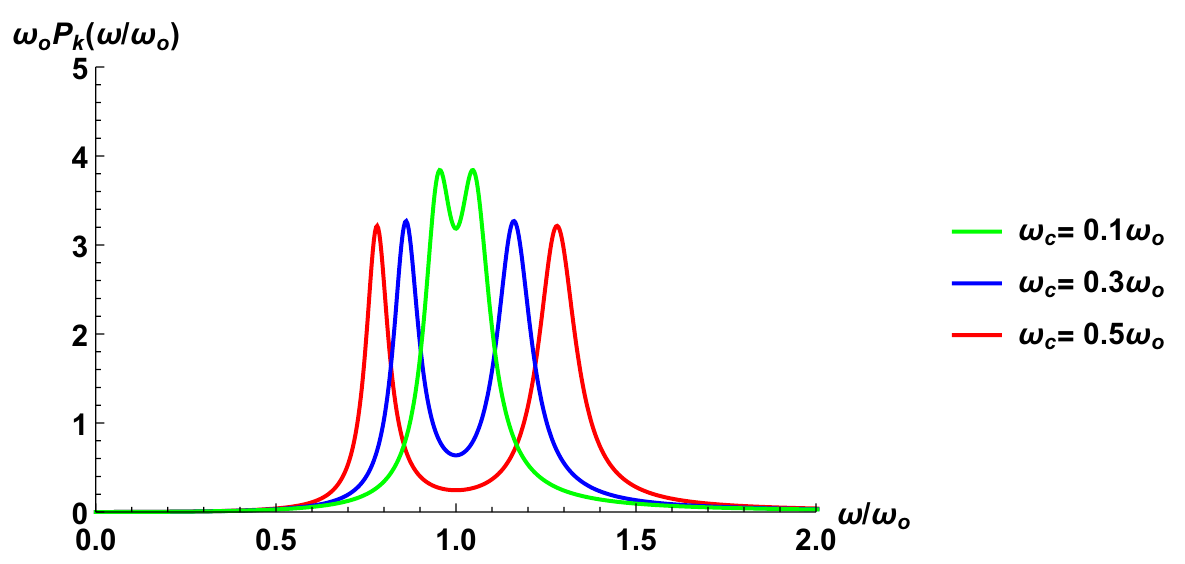}			
		
		\caption{\footnotesize Variation of dimensionless kinetic energy distribution \(\omega_0 P_k(\omega/\omega_0)\) versus rescaled thermostat frequency \(\omega/\omega_0\) for \(\gamma_0/\omega_0 = 0.1\) and different values of cyclotron frequency.}
	
\end{figure}

\begin{figure}
	\centering
		\includegraphics[width=4.3in]{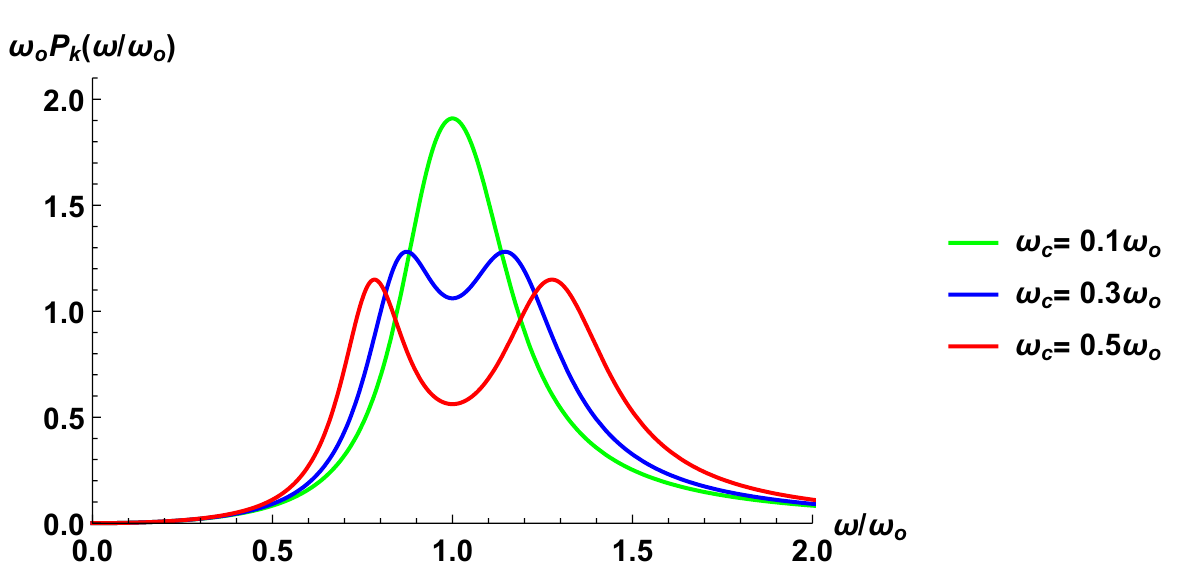}			
		
		\caption{\footnotesize Variation of dimensionless kinetic energy distribution \(\omega_0 P_k(\omega/\omega_0)\) versus rescaled thermostat frequency \(\omega/\omega_0\) for \(\gamma_0/\omega_0 = 0.3\) and different values of cyclotron frequency.}
	
\end{figure}

Let us now comment on the effect of damping strength parameter\footnote{The parameter \(\gamma_0\) shall be interchangeably referred to as damping strength, dissipation strength, coupling strength, etc.} \(\gamma_0\). From figures-(3) and (4), it becomes clear that an increase in the damping strength leads to flattening of the curves, i.e. for larger strengths of coupling between the dissipative oscillator and the thermostat, a wider window of thermostat oscillators contribute to the mean kinetic energy of the system. On the other hand, as the coupling strength gets weaker, the curves sharpen out, and therefore, bath oscillators whose frequencies are around the peaks are able to make a significant contribution to the mean kinetic energy of the system. In fact, from figure-(4), we notice that stronger values of dissipation strength may even lead to coalescence of the peaks. This is interesting, because increasing the damping strength seems to act against the effect of the applied magnetic field, because an increase in the latter value leads to a more pronounced pair of peaks. This points out towards the effect of decoherence, induced by dissipation \cite{dec}, the level of which increases with the damping strength. On the other hand, an applied magnetic field is responsible for coherent dynamics (leading to Landau diamagnetism), and therefore, the magnetic field and the dissipation strength compete with each other in this respect. \\

\begin{figure}
	\centering
		\includegraphics[width=4.3in]{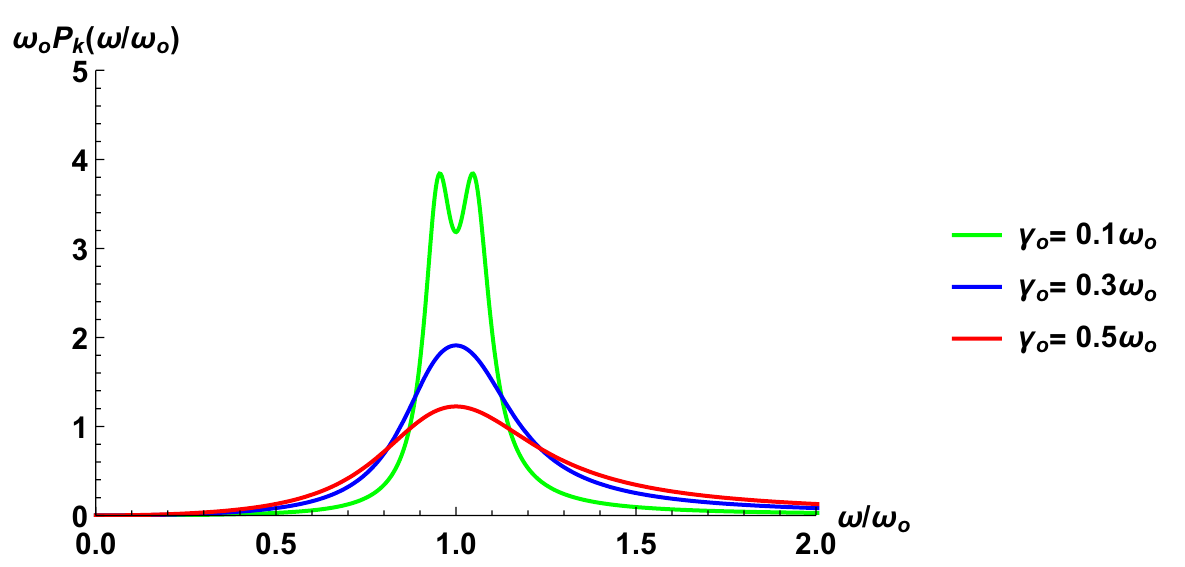}			
		
		\caption{\footnotesize Variation of dimensionless kinetic energy distribution \(\omega_0 P_k(\omega/\omega_0)\) versus rescaled thermostat frequency \(\omega/\omega_0\) for \(\omega_c/\omega_0 = 0.1\) and different values of damping strength.}
	
\end{figure}

\begin{figure}
	\centering
		\includegraphics[width=4.3in]{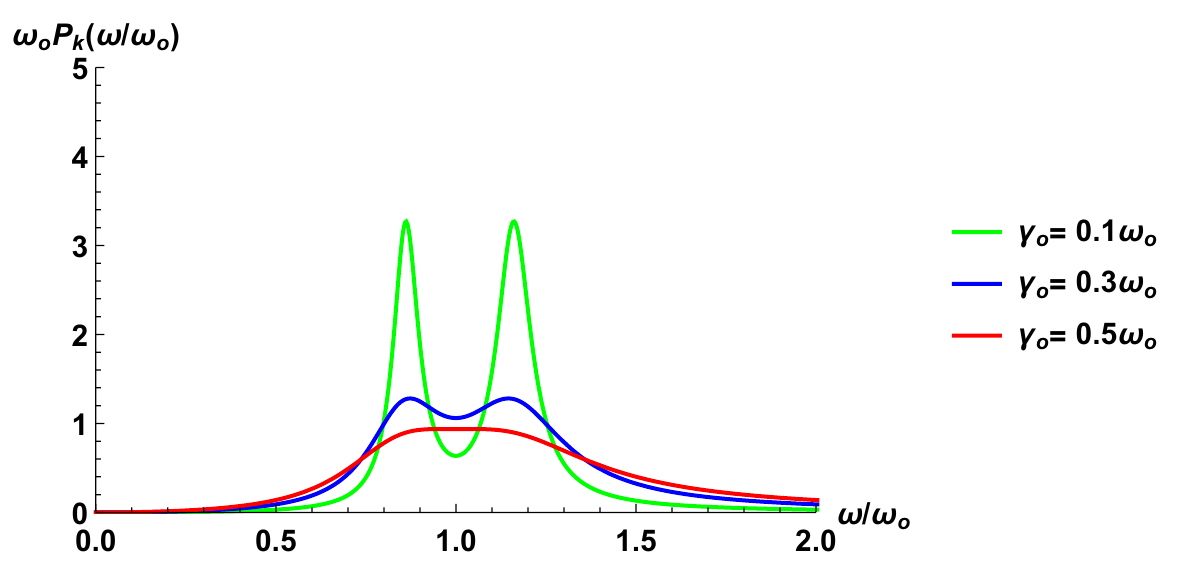}			
		
		\caption{\footnotesize Variation of dimensionless kinetic energy distribution \(\omega_0 P_k(\omega/\omega_0)\) versus rescaled thermostat frequency \(\omega/\omega_0\) for \(\omega_c/\omega_0 = 0.3\) and different values of damping strength.}
	
\end{figure}

The location of the peaks and their functional dependence on the various system control parameters \(\gamma_0\), \(\omega_0\), and \(\omega_c\) is hard to obtain analytically. However, we may consider the weak-coupling limit, for which Eq. (\ref{pkgenexp}) which appears inside the integral given in Eq. (\ref{Ekpart}) effectively becomes
\begin{equation}
P_k(\omega)\bigg|_{\gamma_0 \rightarrow 0} \approx \frac{1}{2} \Big[ \delta(\omega - \omega_1) + \delta(\omega-\omega_2)\Big],
\end{equation}
where \(\omega_{1,2}\) are the solutions of the equations \(\omega^2-\omega_0^2 \pm \omega\omega_c = 0\) for \(\omega \geq 0\), and are given by
\begin{equation}\label{omega12}
\omega_{1,2} = \frac{1}{2} (\sqrt{4\omega_0^2 + \omega_c^2} \pm \omega_c ).
\end{equation}
 These are precisely the normal modes of a charged oscillator moving in two dimensions, in the presence of a transverse magnetic field \cite{malay2,abcd4}. This is not surprising, because we have taken \(\gamma_0 \rightarrow 0\), and thus Eq. (\ref{omega12}) does not depend on \(\gamma_0\). In this limit, the mean kinetic energy of the system becomes
\begin{equation}\label{Ekweak1}
(E_k)_{\gamma_0 \rightarrow 0} \approx  \frac{\hbar \omega_1}{4} \coth \bigg(\frac{ \hbar \omega_1}{2 k_B T}\bigg) +  \frac{\hbar \omega_2}{4} \coth \bigg(\frac{ \hbar \omega_2}{2 k_B T}\bigg).
\end{equation}
The appearance of temperature \(T\) in the above equation reminds us that the system is still coupled to the thermostat, although we have just taken the strength of the coupling to be small. This is precisely an assumption used in elementary statistical mechanics textbooks, to derive the canonical partition function \cite{book}. In a sense, therefore, the microscopic approach based on the quantum Langevin equation, and subsequently deriving the energy from it is more general, with the textbook result [Eq. (\ref{Ekweak1})] emerging as a special case. If further, one puts \(\omega_c = 0\), we have \(\omega_{1,2} = \omega_0\), meaning that
\begin{equation}
(E_k)_{\gamma_0,\omega_c \rightarrow 0} \approx  \frac{\hbar \omega_0}{2} \coth \bigg(\frac{ \hbar \omega_0}{2 k_B T}\bigg),
\end{equation} which is the mean kinetic energy of a two-dimensional isotropic quantum oscillator with frequency \(\omega_0\). Therefore, in a sense, the weak-coupling limit indicates that only those thermostat degrees of freedom which resonate with the frequency of the system (for \(\omega_c \neq 0\), there are two frequencies) are relevant, motivating the rotating-wave approximation \cite{RWA}. We summarize by noting that in general, the probability distribution function \(P_k(\omega)\) is extremely sensitive to control parameters such as damping strength, harmonic trap frequency, and strength of magnetic field. Therefore, it can be easily manipulated in an experimental setting.

\section{Magnetic moment} \label{magneticmomentsec}
We shall now demonstrate a connection between the equilibrium magnetic moment of the dissipative oscillator and the energy partition theorem. The magnetic moment of the oscillator can be computed from the following correlation function \cite{sdg1}:
\begin{eqnarray}
M_z &=& \frac{|e|}{2 c} \langle x(t) \dot{y}(t) - y(t) \dot{x}(t) \rangle \\
 &=&  \frac{|e|}{4 c} {\rm Im} \langle \dot{Z}(t) Z(t)^\dagger + Z(t)^\dagger \dot{Z}(t) \rangle,
\end{eqnarray} where \(Z(t)\) and \(\dot{Z}(t)\) are given by Eqs. (\ref{Z}) and (\ref{dotZ}). With a few straightforward manipulations, it follows that in the steady state,
\begin{equation}\label{mz}
M_z = -\frac{e \hbar}{4 \pi m c}  \int_{-\infty}^\infty d\omega \omega^2 \coth \bigg(\frac{ \hbar \omega}{2 k_B T} \bigg) [\Phi(\omega) - \Phi(-\omega)].
\end{equation} Thus, upon identifying \(m (\omega) = -\frac{e \hbar}{ 2 m c} \coth (\frac{ \hbar \omega}{2 k_B T} )\) (not to be confused with mass), Eq. (\ref{mz}) corresponds to Eq. (\ref{magneticpartition}) thereby proving theorem-(\ref{theorem2}). One can cast Eq. (\ref{mz}) in a more intuitive form by putting \(\omega \rightarrow -\omega\) in the second term, i.e. in the integral involving \(\Phi ( -\omega)\) so that one gets
\begin{equation}\label{mz1}
M_z = \frac{1}{\pi}  \int_{-\infty}^\infty d\omega \omega^2 m(\omega) \Phi(\omega).
\end{equation} In terms of a function \(\mathcal{P}_m (\omega) := \pi^{-1} \omega^2 \Phi(\omega) \), we have
\begin{equation}\label{mz11}
M_z = \int_{-\infty}^\infty d\omega m(\omega) \mathcal{P}_m (\omega).
\end{equation}
This is manifestly negative due to the fact that \(m(\omega)\) carries an overall negative sign. Equation (\ref{mz11}) is a remarkable result because the magnetic moment of the system is written as a sum over the entire bath spectrum such that \( m(\omega) \mathcal{P}_m (\omega) d \omega\) refers the contribution coming from frequency range \(\omega\) to \(\omega + d\omega\). Surprisingly, \(\mathcal{P}_m(\omega)\) is a probability distribution function because it is normalized [Eq. (\ref{omegaPhinormalize})] and positive definite. In fact, we may also rewrite Eq. (\ref{Ekpart}) as 
\begin{equation}\label{Ekpart11}
E_k = \int_{-\infty}^\infty \mathcal{E}(\omega) \mathcal{P}_m(\omega) d\omega,
\end{equation} meaning that, if one considers the domain \(\omega \in (-\infty, \infty)\), i.e. including negative-phasor portion of the Fourier spectrum, then the mean kinetic energy and equilibrium magnetic moment are described by the same probability distribution function. Thus, there seems to be a connection between the probability distribution functions describing partition of kinetic energy and magnetic moment in dissipative diamagnetism. In fact, the following result holds:

\begin{theorem}\label{theorem3}
Consider a smooth and even function \(F : \mathbb{R} \rightarrow \mathbb{R}\) on the real line. It can be represented as \(F(\omega)\), where \(\omega \in (-\infty, \infty)\) and it satisfies \(F(-\omega) = F(\omega)\). Then, its mean with respect to the probability distribution \(\mathcal{P}_m(\omega)\), denoted by \(\langle \cdot \rangle_{\mathcal{P}_m}\) is given by
\begin{equation}
\langle F (\omega) \rangle_{\mathcal{P}_m} = \sum_{n=0}^\infty a_{2n} \langle \omega^{2n} \rangle_{P_k},
\end{equation} where \(\langle \cdot \rangle_{P_k}\) denotes averaging with respect to the distribution function \(P_k(\omega)\) over the interval \(\omega \in [0,\infty)\), and \(\{a_{2n}\}\) are the even coefficients in the Taylor expansion of \(F(\omega)\) about the origin. 
\end{theorem}

\textit{Proof -}  The above result can be proven by straightforward manipulations. Since \(F(\omega)\) is smooth, we may consider its Taylor expansion about \(\omega = 0\): 
\begin{equation}
F(\omega) = \sum_{n=0}^\infty (a_{2n} \omega^{2n} + a_{2n + 1} \omega^{2n+1}),
\end{equation} where \(\{a_{2n}\}\) and \(\{a_{2n+1}\}\) are all real numbers. Since the function is even, we have \(a_{2n+1} = 0\) for all \(n\). Then, the average of \(F(\omega)\), with respect to the distribution \(\mathcal{P}_m(\omega)\) is 
\begin{equation}\label{Favg1}
\langle F (\omega) \rangle_{\mathcal{P}_m} = \sum_{n=0}^\infty a_{2n} \int_{-\infty}^\infty   \omega^{2n}  \mathcal{P}_m (\omega) d\omega =  \sum_{n=0}^\infty \frac{a_{2n}}{\pi} \int_{-\infty}^\infty   \omega^{2n+2} \Phi(\omega) d\omega.
\end{equation}
Consider the integral
\begin{equation}\label{Iint}
I = \int_{-\infty}^\infty   \omega^{2n+2} \Phi(\omega) d\omega.
\end{equation} Putting \(\omega \rightarrow -\omega\), and adding to Eq. (\ref{Iint}), we have
\begin{equation}
I = \frac{1}{2} \int_{-\infty}^\infty   \omega^{2n+2} [\Phi(\omega) +\Phi(-\omega)]d\omega =  \int_{0}^\infty   \omega^{2n+2} [\Phi(\omega) +\Phi(-\omega)]d\omega.
\end{equation}
Substituting this back into Eq. (\ref{Favg1}) straightforwardly gives 
\begin{equation}
\langle F (\omega) \rangle_{\mathcal{P}_m}  =  \sum_{n=0}^\infty \frac{a_{2n+2}}{\pi} \int_{0}^\infty   \omega^{2n} [\Phi(\omega) +\Phi(-\omega)] d\omega = \sum_{n=0}^\infty a_{2n} \int_0^\infty \omega^{2n} P_k (\omega) d\omega,
\end{equation} thereby proving the result. \\

Some discussion is in order. One may notice that the magnetic moment arises from one of the terms of the kinetic energy expression appearing in the Hamiltonian [Eq. (\ref{H})] i.e. from $\frac{e^2B^2}{8mc^2}(x^2+y^2)$ \cite{ashcroft}. As a result, both of them are governed by the same probability density. This is somewhat intuitive because the magnetic moment originates from the term that is linked with the canonical momentum of the motion of a charged particle in an external uniform magnetic field which is ideally used to model our system of interest. However, there is one important difference between the quantum counterpart of the energy equipartition theorem and Eq. (\ref{mz11}). In Eq. (\ref{Ekpart}) or (\ref{Ekpart11}), the quantity \(\mathcal{E}_k(\omega)\) refers to the mean energy of an individual bath oscillator in frequency range \(\omega\) to \(\omega + d\omega\). On the other hand, in Eq. (\ref{mz11}), the quantity \(m(\omega)\), bears no such interpretation because the bath oscillators are electrically neutral and cannot possess a magnetic moment! Nevertheless, Eq. (\ref{mz11}) offers a new perspective to dissipative diamagnetism, that the magnetic moment at equilibrium can be expressed as a sum taken over an appropriate probability distribution function and indicates the connection between the quantum counterpart of energy equipartition theorem and dissipative diamagnetism.\\

It is imperative to check whether Eq. (\ref{mz11}) gives \(M_z = 0\) for zero external field, i.e. \(\omega_c = 0\). Let us first note that from Eqs. (\ref{mudef}) and (\ref{Jdef}), for any dissipation function \(\mu(t)\), the real and imaginary parts of its Fourier transform are respectively even and odd functions in \(\omega\). Then, putting \(\omega_c = 0\) makes \(\Phi (\omega)\) (hence, \(\mathcal{P}_m (\omega)\)) an even function making Eq. (\ref{mz11}) vanish because \(m(\omega)\) is odd. On the other hand, Eq. (\ref{kineticenergysteadystate1}) is non-zero (as expected) since \(\mathcal{E}_k (\omega)\) is an even function. Another interesting limit is the classical limit, i.e. \(\hbar \rightarrow 0\) for which
\begin{equation}
\frac{ \hbar \omega}{2 k_B T} \coth \bigg( \frac{\hbar \omega}{2 k_B T} \bigg) \rightarrow 1.
\end{equation} Therefore, Eq. (\ref{mz1}) gives
\begin{equation}\label{mz111}
M_z = -\frac{e k_B T}{ m c \pi}  \int_{-\infty}^\infty d\omega \omega \Phi(\omega) .
\end{equation} This integral can be evaluated for specific choices of parameters. It may be checked that the final answer is vanishingly small, consistent with the Bohr-van Leeuwen theorem. We conclude this section by remarking that although for simplicity, we shall be choosing Ohmic dissipation in sections (\ref{superstatsection}) and (\ref{seriessection}), the results presented in this section are completely general in the sense that they hold for any arbitrary dissipation mechanism satisfying some basic requirements discussed in \cite{ford1988}.

\section{Superstatistics of energy and magnetic moment}\label{superstatsection}
In the previous two sections, we have computed the mean kinetic energy as well as magnetic moment of the two-dimensional dissipative oscillator placed in a transverse magnetic field. In this section, we shall discuss these results in the context of superstatistics (see also \cite{superstat}), namely a superposition of two statistics. To get a naive idea about superstatistics, consider a generic system described by the Boltzmann distribution which means that the probability of finding the system in a state with energy \(E_k\) is proportional to \(e^{-\beta E_k}\) where \(\beta\) is the inverse temperature. Typically, the system can have fluctuations about its local inverse temperature. If these fluctuations are associated with a distribution function \(f(\beta)\), then one can define a superstatistical Boltzmann distribution as \cite{SS1,SS2}
\begin{equation}
B(E_k) = \int_0^\infty f(\beta) e^{-\beta E_k} d\beta,
\end{equation} such that the partition function reads \(Z = \sum_k B(E_k)\). Thus, the physical quantities which are computed using the partition function are a result of a two-fold average: the first averaging takes places over temperature fluctuations, whereas the second one involves tracing over all accessible energy levels. In the case of the Brownian oscillator, we too encounter two-fold averaging for computing the quantities of interest such as energy and magnetic moment at equilibrium.\\

We have already computed the mean energy and magnetic moment of the two-dimensional dissipative oscillator in the presence of an external magnetic field, and it was found that both energy and magnetic moment could be expressed as integrals over the bath spectrum. A common theme in these results is that an averaged quantity \(X\) at temperature \(T\) is expressed as
\begin{equation}\label{XT}
X(T) = \int_{-\infty}^\infty x(\omega,T) \mathcal{P}_x(\omega) d\omega,
\end{equation} where \(\mathcal{P}_x(\omega)\) is a temperature independent probability distribution function, i.e. it is both positive definite and normalized:
\begin{equation}
\mathcal{P}_x(\omega) \geq 0, \forall \omega \in (-\infty,\infty), \hspace{5mm}  \int_{-\infty}^\infty \mathcal{P}_x(\omega) d\omega = 1.
\end{equation}
In Eq. (\ref{XT}), \(x(\omega,T)\) is obtained via a suitable averaging over the Gibbs canonical state of the bath at temperature \(T\). For example, in the case of total energy \(x(\omega,T) = \mathcal{E}(\omega, T)\) is the mean total energy of a two-dimensional bath oscillator. This mean is obtained from the Gibbs ensemble describing the bath as
\begin{equation}
\mathcal{E} (\omega,T) = - \lim_{N \rightarrow \infty} \frac{1}{N} \frac{\partial}{\partial \beta}  \ln Z_{\rm bath} = \lim_{N \rightarrow \infty} \frac{1}{N}  \frac{{\rm Tr}( E \rho_{\rm bath})}{{\rm Tr}(\rho_{\rm bath})},
\end{equation} where \(N\) is the number of independent bath oscillators (infinite in number to ensure thermodynamic equilibrium), \(\rho_{\rm bath}\) is the density matrix of the bath at temperature \(T\), \(Z_{\rm bath} = {\rm Tr}(\rho_{\rm bath})\) is the associated canonical partition function, and \(\beta = 1/T\). Although we are considering the total (kinetic + potential) energy above, the kinetic and potential energy cases may also be discussed separately in the same spirit as above\footnote{Although we shall explicitly consider the kinetic energy, one can analogously analyze the potential energy using the same method.}. \\

For the case of magnetic moment, we have \(x(\omega,T) = m(\omega,T) = - \mu_B \coth \big(\frac{\hbar \omega}{2 k_B T}\big)\) which is the thermal Bohr magneton. It may be obtained from the partition function of the bath as
\begin{equation}\label{thermbohr}
m(\omega,T) = \frac{e}{mc} \lim_{N \rightarrow \infty} \frac{1}{N} \frac{1}{\beta} \frac{\partial}{\partial \omega} \ln Z_{\rm bath}.
\end{equation}
Therefore, we note that both for energy and magnetic moment, associated quantities for the system (the Brownian oscillator) are obtained as a two-fold average where the first averaging is performed over the canonical state of the bath, whereas the second one takes place over the entire bath spectrum. This motivates the use of superstatistics \cite{superstat,SS1,SS2}. Below, we consider kinetic energy and magnetic moment separately.

\subsection{Superstatistics of kinetic energy}
Since the expression for kinetic energy involves a double averaging, reminiscent of superstatistics, we can re-express this as an average taken over the kinetic energy of the bath oscillators \cite{superstat}. Let us begin by noting that \(\mathcal{E}_k(\omega)\) is given by
\begin{equation}\label{Eomegarelation}
\mathcal{E}_k ( \omega) = \frac{\hbar \omega}{2} \coth \bigg(\frac{\hbar \omega}{2 k_B T}\bigg) .
\end{equation} We may invert this relation to find \(\omega = \omega(\mathcal{E}_k)\). Thus, the quantity \(\mathcal{E}_k (\omega) P_k (\omega) d\omega\) can be rewritten as an integral over \(\mathcal{E}_k\) by a change of variables from \(\omega\) to \(\mathcal{E}_k\) as
\begin{equation}\label{vvv}
E_k = \int_{0}^\infty \mathcal{E}_k(\omega) P_k (\omega) d\omega = \int_{k_B T}^\infty \mathcal{E}_k f_k (\mathcal{E}_k,T) d\mathcal{E}_k,
\end{equation} where
\begin{equation}\label{vvvv}
f_k(\mathcal{E}_k,T) = P_k(\omega(\mathcal{E}_k,T)) \frac{d \omega}{d\mathcal{E}_k}.
\end{equation} In Eqs. (\ref{vvv}) and (\ref{vvvv}), one uses the relation \(\omega = \omega(\mathcal{E}_k)\) to express the right-hand side as a function of \(\mathcal{E}_k\) (and \(T\)), while \(d\omega/d\mathcal{E}_k\) can be found as the reciprocal of the equation
\begin{equation}
\frac{d\mathcal{E}_k}{d\omega} = \frac{\hbar}{2}\Bigg[ \coth \bigg(\frac{\hbar \omega}{2 k_B T}\bigg) - \bigg(\frac{\hbar \omega}{2 k_B T}\bigg) {\rm cosech}^2 \bigg(\frac{\hbar \omega}{2 k_B T}\bigg) \Bigg].
\end{equation}
It is noteworthy that
\begin{equation}
\int_{k_B T}^\infty f_k (\mathcal{E}_k,T) d\mathcal{E}_k = \int_{0}^\infty  P_k (\omega) d\omega = 1,
\end{equation} implying that \(f_k(\mathcal{E}_k,T)\) is a normalized distribution function in the energy representation. One may notice the following mathematical properties (see also \cite{superstat}): (i) $f_k(\mathcal{E}_k,T)$ can be defined in the interval $[\mathcal{E}_0,\infty)$ where $\mathcal{E}_0=k_BT$; (ii) $f_k(\mathcal{E}_k,T)\rightarrow \infty$ when $\mathcal{E}\rightarrow \mathcal{E}_0$, and $f_k(\mathcal{E}_k,T)\rightarrow 0 $ for $\mathcal{E}\rightarrow \infty $; (iii) $\int_{\mathcal{E}_0}^\infty f_k (\mathcal{E}_k,T) d\mathcal{E}_k =1$. One should also note that unlike \(P_k(\omega)\), the distribution function \(f_k(\mathcal{E}_k,T)\) is sensitive to temperature, in addition to its dependence on system and bath parameters. Thus, the two-fold average can be expressed as an average taken over the kinetic energies of the bath oscillators in the sense that the quantity \(\mathcal{E}_k f_k (\mathcal{E}_k,T) d\mathcal{E}_k\) represents the contribution to the dissipative oscillator's mean kinetic energy from the interval \(\mathcal{E}_k\) to \(\mathcal{E}_k + d\mathcal{E}_k\). Since solving Eq. (\ref{Eomegarelation}) involves solving for a transcendental equation, we resort to finding \(f_k(\mathcal{E}_k,T) \) numerically. In figure-(5), we have plotted the new distribution function \(f_k(\mathcal{E}_k,T) \), as a function of \(\mathcal{E}_k\) for three different values of the cyclotron frequency. 

\begin{figure}
	\centering
		\includegraphics[width=4.3in]{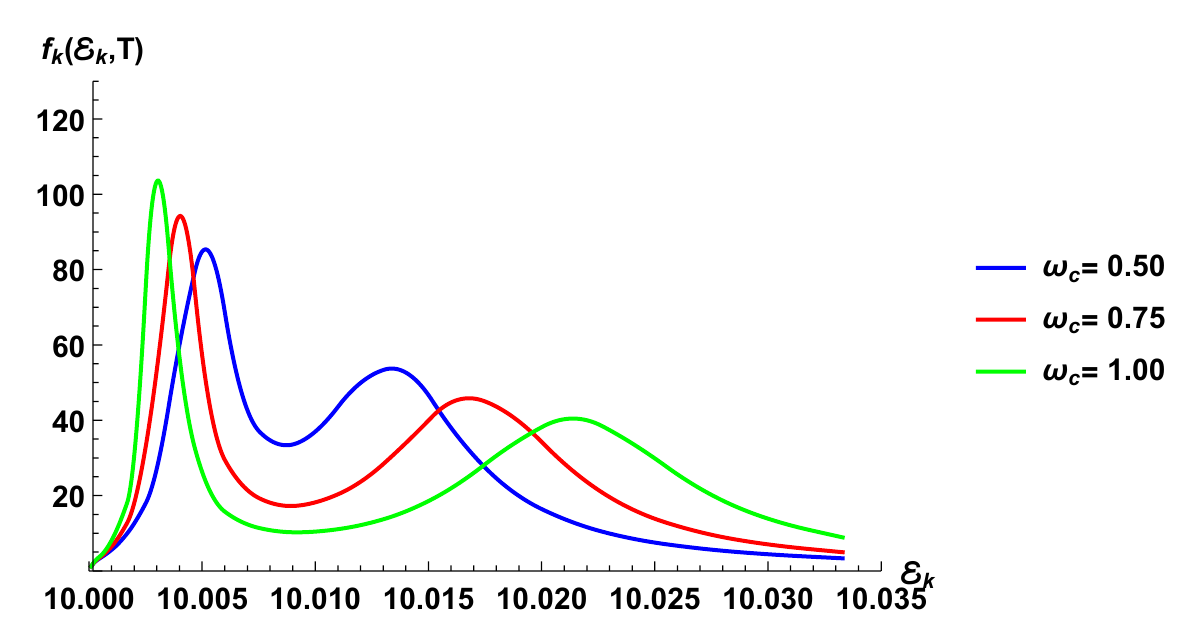}		
		
		\caption{\footnotesize Variation of the distribution function \(f_k(\mathcal{E}_k,T)\) versus \(\mathcal{E}_k\) for Ohmic dissipation with \(\hbar = \omega_0 = 1\), \(\gamma_0 = 0.3\), and different values of cyclotron frequency.}
	
\end{figure}

\subsection{Superstatistics of magnetic moment}\label{ssmmsec}
Let us now consider the superstatistics of the magnetic moment. To begin with, let us first note that the equilibrium magnetic moment of the two-dimensional dissipative charged oscillator is given by Eq. (\ref{mz11}), i.e. \(M_z = \int_{-\infty}^\infty m(\omega) \mathcal{P}_m (\omega) d\omega\) where \(m(\omega)\) is the thermal Bohr magneton [Eq. (\ref{thermbohr})] and \(\mathcal{P}_m(\omega) = \pi^{-1} \omega^2 \Phi(\omega)\) is a probability distribution function in \(\omega \in (-\infty,\infty)\). Following the analysis in the previous section, we define a new distribution function \(f_m (m,T)\) as
\begin{equation}
f_m(m,T) := \mathcal{P}_m(\omega(m,T)) \frac{d \omega}{dm},
\end{equation} where \(\omega = \omega(m)\) is found by inverting the relation \(m = m(\omega)\) [Eq. (\ref{thermbohr})] (which also contains \(T\)). In terms of this new function \(f_m(m,T)\), the equilibrium magnetic moment of the dissipative oscillator reads
\begin{equation}
M_z = \int_{-\infty}^\infty m f_m(m,T) dm.
\end{equation}
One should note that unlike the function \(f_k(\mathcal{E}_k,T)\) discussed in the previous subsection which is defined over \(\mathcal{E}_k \in (k_BT,\infty)\), the function \(f_m(m,T)\) is defined from \(m \in ( -\infty,\infty)\). However, the distribution function \(f_m (m,T)\) is not positive definite due to the fact that \(d\omega/dm\) is not positive definite. It has been plotted in figure-(6) for different values of the magnetic field parameter \(\omega_c\). Our results show the existence of an interesting picture, which can be considered as complementary to the
existing ones. We show  that the equilibrium
state of the dissipative magnetic system is characterized by a wide magnetic moment distribution. The areas enclosed by the positive and negative wings are in general unequal resulting in a net negative magnetic moment (see subsection (\ref{magneticmomentseries})). However, it may be checked that in the high-temperature
limit ($k_B T >> \hbar \omega_0$), the positive and negative contributions almost cancel each other. This is consistent with the Bohr-van Leeuwen theorem. On the other hand,
at low temperatures, the contributions of the positive and negative segments of the distribution do not exactly cancel, giving a net non-zero magnetic moment. One may speculate that the existence of the two wings of the distribution is intimately related to the initial ideas of Bohr and van Leeuwen concerning opposite contributions of the `bulk' rotating currents (corresponding to the angular momentum part $L_z$) and `surface' current contributions
 to the total magnetic moment distribution (see for example \cite{peirls} about the qualitative description
of these two contributions).

\begin{figure}
	\centering
		\includegraphics[width=4.5in]{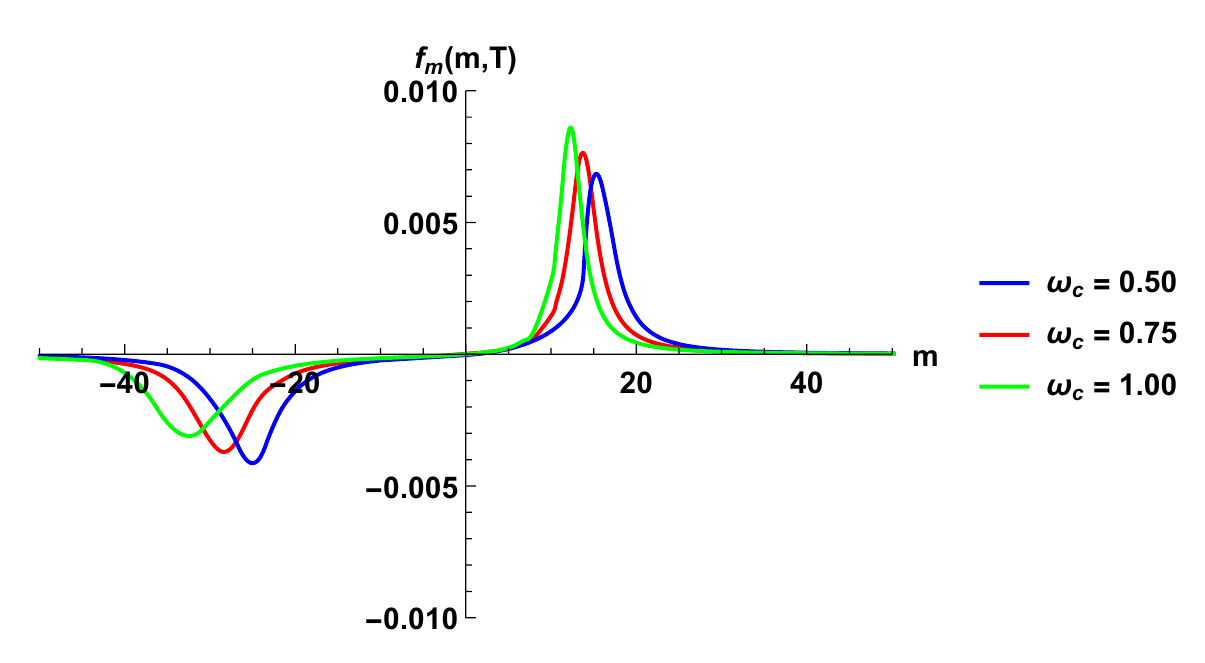}			
		
		\caption{\footnotesize Variation of the distribution function \(f_m(m,T)\) versus \(m\) for Ohmic dissipation with \(\hbar = \omega_0 = 1\), \(\gamma_0 = 0.3\), and different values of cyclotron frequency.}
	
\end{figure}

\section{Equivalence between the Einstein and Gibbs methods}\label{seriessection}
In this section, we are going to demonstrate the equivalence of two distinct approaches to the statistical mechanics
of dissipative quantum systems, viz., the ensemble approach of Gibbs and the equation-of-motion
approach due to Einstein utilizing the paradigmatic model of dissipative diamagnetism. For definiteness, we shall consider Ohmic dissipation, where \(\tilde{\gamma}(\omega) = \gamma_0\). Let us begin with magnetic moment.

 \subsection{Magnetic moment} \label{magneticmomentseries}
One can manipulate Eq. (\ref{mz}) as
\begin{widetext}
\begin{eqnarray}\label{mag2}
M_z&=&-\frac{e}{\pi mc\beta}\int_{-\infty}^{\infty}d\omega  \Big[\Big(\frac{\beta\hbar\omega}{2}\Big)\coth\Big(\frac{\beta\hbar\omega}{2}\Big)\omega[\Phi(\omega)-\Phi(-\omega)]\Big]\nonumber \\
&=&-\frac{e}{\pi mc\beta}\int_{-\infty}^{\infty}d\omega\Big(\frac{\beta\hbar\omega}{2}\Big)\coth\Big(\frac{\beta\hbar\omega}{2}\Big) {\rm Im} \Big[\chi(\omega)-\chi(-\omega)\Big]\nonumber \\
&=&-\frac{e}{\pi mc\beta}{\rm Im} \sum_{n=1}^{\infty}\int_{-\infty}^{\infty}d\omega  \Big[\frac{\omega}{\omega+i\nu_n}+\frac{\omega}{\omega-i\nu_n}\Big]\Big[\chi(\omega)-\chi(-\omega)\Big].
\end{eqnarray}
Here we have employed the summation formula: $x \coth (x) = 1+ 2\sum_{n=1}^{\infty} (x^2)/[(x^2)+(n\pi)^2]$, where $x$ is in general complex, and we use the fact that the term unity in the above formula multiplies to ${\rm Im}[\chi(\omega) - \chi (-\omega)]$ in Eq. (\ref{mag2}) and thus integrates out to zero. Here $\chi(\omega)=\frac{1}{[(\omega^2-\omega_0^2+\omega\omega_c)-i\gamma_0\omega]}$. The term  $1/(\omega+i\nu_n)$ has a pole at $\omega = -i\nu_n$ in the lower half of the complex plane and thus contributes to the first term (also lying in the lower half-plane) within the third bracket  parentheses of the third line of Eq. (\ref{mag2}). Similarly, the pole at $\omega = +i \nu_n$ in the upper half-plane contributes to the second term (lying in the upper half-plane) within the third bracket parentheses of the third line of Eq. (\ref{mag2}). Hence, after performing the contour integration we obtain
\begin{eqnarray}\label{mag22}
M_z &=& -2\frac{e}{ mc\beta}\sum_{n=1}^{\infty} \frac{\nu_n^2\omega_c}{[\nu_n^2+\omega_0^2+\gamma_0\nu_n]^2+(\nu_n\omega_c)^2},
\end{eqnarray}
\end{widetext}
where $\nu_n=\frac{2\pi n}{\hbar\beta}$ with $n= 0, 1,2, \cdots $ are the (bosonic) Matsubara frequencies. Our final expression [Eq. (\ref{mag22})] matches with Eq. (55) of \cite{PRE79}, as the cut-off $\omega_D$ goes to infinity. Eq. (\ref{mag22}) has been plotted in figures-(7)-(9). Let us make a few remarks about Eq. (\ref{mag22}). Usually, equilibrium results are independent of dissipation parameters that arise from weak system-bath interaction \cite{sdgpuri,gsagarwal}. But, Eq. (\ref{mag22}) suggests a non-trivial role played by \(\gamma_0\) in the context of dissipative diamagnetism. Thus, the dissipative magnetic moment is a rare equilibrium property that is characterized by the dissipation parameter $\gamma_0$ which determines the dissipative dynamics of the open quantum system. One should also note that in the context of contemporary and relevant mesoscopic structures, as one increases the dissipation, the level of decoherence increases. Although Landau diamagnetism as expressed in Eq. (\ref{landauanswer}) (below) originates from the coherent cyclotron motion of the electrons, dissipative diamagnetism, captured by Eq. (\ref{mag22}), intrinsically takes into account the effects of decoherence, due to quantum dissipation \cite{dec}. To illustrate this point, we plot in figure-(7), the dissipative magnetic
moment versus the rescaled dissipation parameter $\gamma_0/\omega_0$. One can observe that as one increases dissipation, the concomitant decoherence enhances and this leads to
decrease in the magnitude of the magnetic moment, as if the Bohr-van Leeuwen theorem is re-established. This is an example of quantum–classical
crossover due to environment-induced decoherence, incorporated by the expression given in Eq. (\ref{mag22}). In figure-(8), we have also established the interplay between the cyclotron parameter $\omega_c/\omega_0$ related to coherence, and the dissipation parameter $\gamma_0/\omega_0$ related to decoherence. One may observe that when the decoherence/dissipation parameter is small, the magnetic moment reaches its saturation value rapidly. On the other hand, an enhanced dissipation parameter acts against coherence and tends to prevent the system from reaching saturation magnetization.\\

 Furthermore, the ratio \(\hbar \omega_0/k_B T\) has significant control over diamagnetism. One can clearly see from figure-(9) that as one increases the value of \(\hbar \omega_0/k_BT\), diamagnetism is more pronounced, indicating towards the fact that diamagnetism is inherently quantum mechanical. For the limit \(k_B T >> \hbar \omega_0\), the magnetic moment becomes negligible, consistent with the Bohr-van Leeuwen theorem. This may also be observed in figure-(9) where, for \(\hbar \omega_0/k_B T \rightarrow 0\), one has \(M_z \rightarrow 0\) while as one increases \(\hbar \omega_0/k_B T\), the magnetic moment increases and then saturates to a maximum value. \\
\begin{figure}
	\centering
		\includegraphics[width=4.3in]{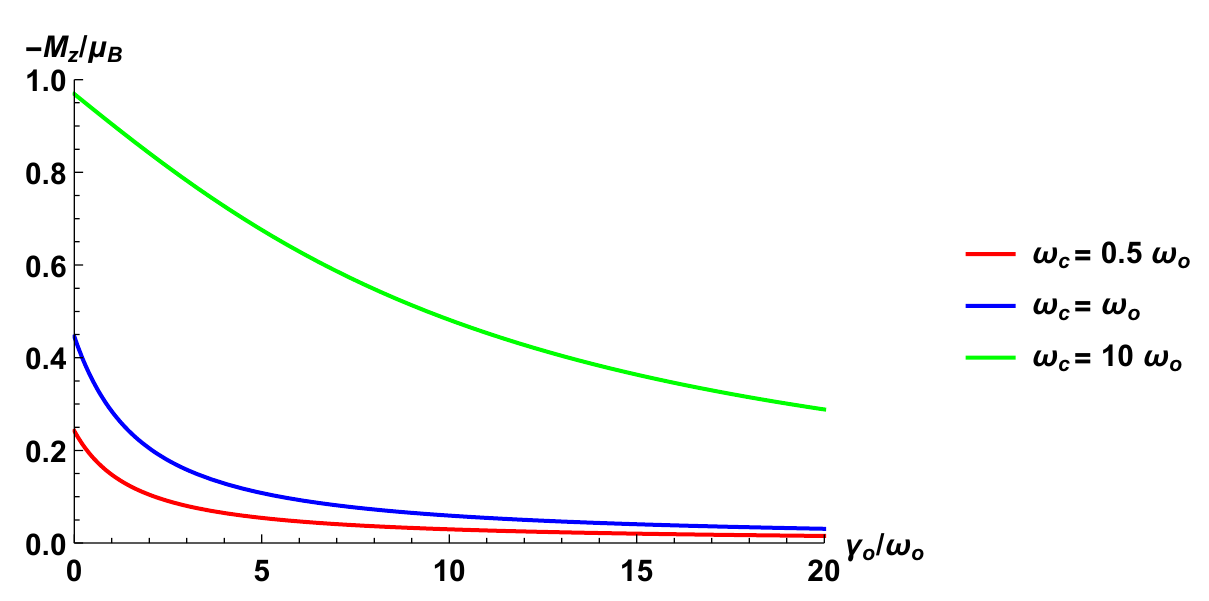}\label{magplot1}		
		
		\caption{\footnotesize Plot of dimensionless magnetic moment \(-M_z/\mu_B\) as a function of rescaled dissipation strength \(\gamma_0/\omega_0\) for different values of rescaled cyclotron frequency \(\gamma_0/\omega_0\) at fixed temperature \(T/ \omega_0=0.1\) (in units \(\hbar = k_B =1\)).}
	
\end{figure}
\begin{figure}
	\centering
		\includegraphics[width=4.3in]{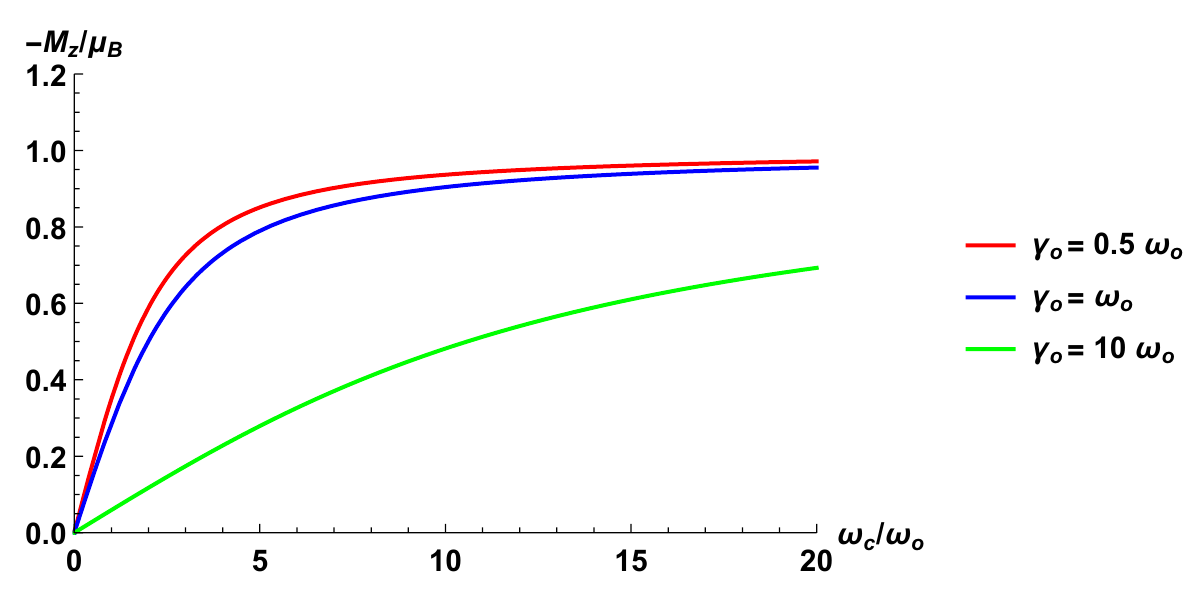}\label{magplot2}		
		
		\caption{\footnotesize Plot of dimensionless magnetic moment  \(-M_z/\mu_B\) as a function of rescaled cyclotron frequency \(\omega_c/\omega_0\) for different values of the ratio \( \gamma_0/\omega_0\) at fixed temperature \(T/ \omega_0=0.1\) (in units \(\hbar = k_B =1\)).}
	
\end{figure}

\begin{figure}
	\centering
		\includegraphics[width=4.3in]{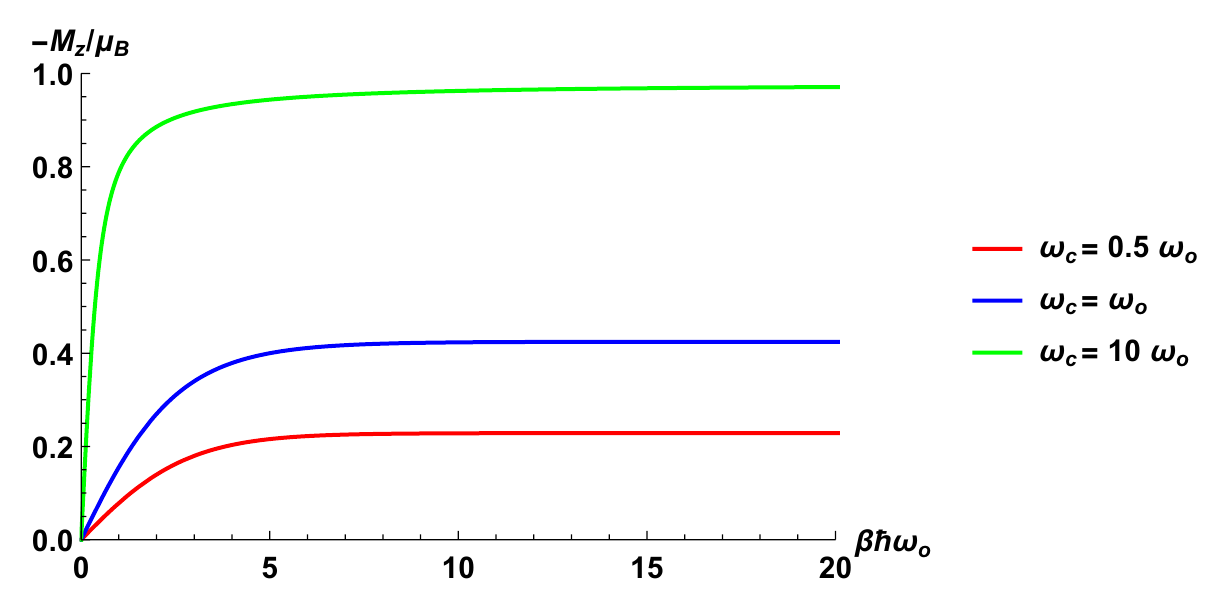}	
\caption{\footnotesize Plot of dimensionless magnetic moment  $-M_z/\mu_B$ as a function of $\hbar \omega_0/k_B T$ for different values of rescaled cyclotron frequency $\omega_c/\omega_0$ and $\gamma_0/\omega_0= 0.1$.}
\end{figure}

Now, we may also consider $\gamma_0 =0 $ in Eq. (\ref{mag22}), for which we obtain
\begin{equation}\label{dissipationzero}
M_z=-\frac{2B}{\beta}\Big(\frac{e}{mc}\Big)^2\sum_{n=1}^{\infty}\frac{\nu_n^2}{(\nu_n^2+\omega_0^2)^2+(\nu_n\omega_c)^2},
\end{equation}
which exactly matches with Eq. (35) of \cite{PRE79} where the latter is calculated from the partition function based on the Gibbs method. Now, if we switch off the harmonic trap by putting $\omega_0 \rightarrow 0$, we can recover famous Landau diamagnetism result:
\begin{eqnarray}
M_z&=&-\frac{2}{B\beta}\sum_{n=1}^{\infty}\frac{\omega_c^2}{(\nu_n^2+\omega_c^2)} \nonumber \\
&=& \frac{e\hbar}{2mc}\Big[\frac{2}{\beta\hbar\omega_c}-\coth\Big(\frac{\beta\hbar\omega_c}{2}\Big)\Big]. \label{landauanswer}
\end{eqnarray}
It should be specially pointed out that the limits \(t \rightarrow \infty\) and \(\omega_0 \rightarrow 0\) do not commute. For obtaining the above result, we have used \(t \rightarrow \infty\) in Eq. (\ref{mz}), followed by \(\omega_0 \rightarrow 0\) in Eq. (\ref{dissipationzero}). Reversing the order of these limits gives a different answer, which is only a part of Landau's result obtained above \cite{sdg1}.

\subsection{Kinetic and potential energies}\label{KEPE}
We shall now express the mean potential and kinetic energies of the dissipative oscillator as an infinite series involving the Matsubara frequencies $\nu_n$. In the process, we shall prove the normalization of \(P_k (\omega)\) defined in section (\ref{mainresultsection}). Following the solution of the quantum Langevin equation given in section (\ref{mainresultsection}), the mean potential energy in the steady state is computed to be
 \begin{eqnarray}\label{pot1}
E_p &=& \lim_{t \rightarrow \infty} \frac{m \omega_0^2}{4} \langle Z(t) Z^\dagger(t) + {\rm c.c}\rangle \nonumber \\
&=&  \frac{\omega_0^2}{2\pi}\int_{-\infty}^{\infty}d\omega \hbar \omega \coth\Big(\frac{\beta\hbar\omega}{2}\Big)\Phi(\omega) \nonumber \\
&=&\frac{\omega_0^2}{\pi\beta}\int_{-\infty}^{\infty}d\omega\Big[1+2\sum_{n=1}^{\infty}\frac{\omega^2}{\omega^2+\nu_n^2}\Big]\Phi(\omega).
\end{eqnarray}
The first term (\(n = 0\)) above can be understood to be the classical result, whereas the subsequent terms (\(n = 1,2,3, \cdots\)) are quantum corrections. Let us consider the $n=0$ term i.e. the term outside the summation in the second line of Eq. (\ref{pot1}). We can rewrite it as
\begin{eqnarray}
({E_p})_{n=0}&=&-\frac{\omega_0^2}{\pi\beta}\int_{-\infty}^{\infty}d\omega \frac{1}{\omega}{\rm Im}\Big[\omega^2-\omega_0^2-\omega\omega_c+i\gamma_0\omega\Big]^{-1}\nonumber \\
&=&\frac{1}{\beta}, \label{norm1}
\end{eqnarray}
where picking up the contribution of the pole at $\omega =0$ provides us the final result. One other way of justifying this contribution (i.e. $\frac{1}{\beta}$) is that as the temperature goes to infinity (in the classical limit) $\coth(\beta\hbar\omega/2)$ goes to $\frac{2}{\beta\hbar\omega}$ and hence, Eq. (\ref{pot1}) reduces to
the $n=0$ term in Eq. (\ref{pot1}). Now, from Eq. (\ref{pot1}) we can rewrite
\begin{widetext}
\begin{eqnarray}
E_p &=& \frac{1}{\beta}+\frac{2\omega_0^2}{\pi \beta}\sum_{n=1}^{\infty}\int_{-\infty}^{\infty}d\omega \omega \Phi(\omega)\frac{\omega}{\omega^2+\nu_n^2} \nonumber \\
&=&\frac{1}{\beta}+\frac{2\omega_0^2}{\pi \beta}\sum_{n=1}^{\infty}\int_{-\infty}^{\infty}d\omega\frac{\omega}{\omega^2+\nu_n^2}{\rm Re}\Big[\frac{i}{\omega^2-\omega_0^2 -\omega\omega_c+i\gamma_0\omega}\Big]\nonumber\\
&=&\frac{1}{\beta}+\frac{2\omega_0^2}{ \beta}\sum_{n=1}^{\infty}{\rm Re}\Big[\frac{1}{\nu_n^2+\omega_0^2+i\nu_n\omega_c+\gamma_0\nu_n}\Big],
\end{eqnarray}
\end{widetext}
where in the last step, we have closed the contour in the upper half-plane and picked up the contribution from the pole at $\omega =i\nu_n$. Finally, we obtain
\begin{equation}\label{pef}
E_p=\frac{1}{\beta}+\frac{2\omega_0^2}{ \beta}\sum_{n=1}^{\infty}\frac{\nu_n^2+\omega_0^2+\gamma_0\nu_n}{(\nu_n^2+\omega_0^2+\gamma_0\nu_n)^2+(\nu_n\omega_c)^2},
\end{equation} which expresses the mean potential energy of the oscillator as an infinite series. \\

Turning now to the calculation of the average kinetic energy, we have from Eq. (\ref{kineticenergysteadystate}), 
\begin{eqnarray}
E_k &=& \frac{1}{2\pi\beta}\int_{-\infty}^{\infty}d\omega \frac{\beta\hbar\omega}{2}\coth\Big(\frac{\beta\hbar\omega}{2}\Big)\omega^2[\Phi(\omega)+\Phi(-\omega)] \nonumber \\
&=&\frac{1}{\beta}+\frac{1}{\pi\beta}\sum_{n=1}^{\infty}\int_{-\infty}^{\infty}d\omega \frac{\omega^2}{\omega^2+\nu_n^2}\omega^2 [\Phi(\omega)+\Phi(-\omega)],
\end{eqnarray}
wherein we have used an argument akin to that of Eq. (\ref{norm1}) to suggest that
\begin{equation}
\frac{1}{2\pi}\int_{-\infty}^{\infty}d\omega \omega^2 [\Phi(\omega) + \Phi(-\omega)] = \frac{1}{\pi}\int_{-\infty}^{\infty}d\omega \omega^2 \Phi(\omega) = 1.
\end{equation}
This implies the normalization of \(P_k (\omega)\) defined earlier. Now, we can express $\omega^2 \Phi(\omega)$ and $\omega^2 \Phi(-\omega)$ as
\begin{eqnarray}
\omega^2\Phi(\omega)&=&{\rm Re} \Big[\frac{i\omega}{\gamma_0}\frac{1}{(\omega^2-\omega_0^2-\omega\omega_c)+i\gamma_0\omega}\Big],  \\
\omega^2\Phi(-\omega)&=&{\rm Re}\Big[\frac{-i\omega}{\gamma_0}\frac{1}{(\omega^2-\omega_0^2+\omega\omega_c)-i\gamma_0\omega}\Big],
\end{eqnarray}
\begin{widetext}
which means that we can write
\begin{eqnarray}
E_k =\frac{1}{\beta}&+&\frac{1}{\pi\beta}\sum_{n=1}^{\infty}\int_{-\infty}^{\infty}d\omega \frac{\omega^2}{(\omega+i\nu_n)(\omega-i\nu_n)}\Big[{\rm Re}\Big\lbrace\frac{i\omega}{\gamma_0}\frac{1}{(\omega^2-\omega_0^2-\omega\omega_c)+i\gamma_0 \omega}\Big\rbrace \nonumber \\
&+&{\rm Re} \Big\lbrace\frac{-i\omega}{\gamma_0}\frac{1}{(\omega^2-\omega_0^2+\omega\omega_c)-i\gamma_0\omega}\Big\rbrace\Big].
\end{eqnarray}
\end{widetext}
Finally, picking up the contribution of the pole in the lower half-plane at $\omega = -i\nu_n$, one may obtain
\begin{eqnarray}\label{kef}
E_k =\frac{1}{\beta}+\frac{2}{\beta}\sum_{n=1}^{\infty}\frac{\Big[\gamma_0 \nu_n\Big(\nu_n^2+\omega_0^2+\gamma_0 \nu_n\Big)+(\nu_n\omega_c)^2\Big]}{\Big[\Big(\nu_n^2+\omega_0^2+\gamma_0 \nu_n\Big)^2+(\omega_c\nu_n)^2\Big]}. 
\end{eqnarray}
Combining Eqs. (\ref{pef}) and (\ref{kef}), we can obtain the internal energy of the system as
\begin{equation}\label{int1}
E =\frac{2}{\beta}\Bigg[1+\sum_{n=1}^{\infty}\frac{N(\nu_n)}{D(\nu_n)} \Bigg],
\end{equation}
where the numerator $N(\nu_n)$ and the denominator $D(\nu_n)$ are given by
\begin{eqnarray}\label{int2}
&&N(\nu_n)=\Big(\nu_n^2+\omega_0^2+\gamma_0\nu_n\Big)\Big(\omega_0^2+\gamma_0\nu_n\Big)+(\omega_c\nu_n)^2 , \nonumber \\
&& D(\nu_n)=\Big[\Big(\nu_n^2+\omega_0^2+\gamma_0\nu_n\Big)^2+(\omega_c\nu_n)^2\Big].
\end{eqnarray}
At this present outset, we can compare our result obtained using the Einstein approach with that obtained via the more conventional Gibbs thermodynamics method. From Eq. (42) of \cite{sdg2} we can write
\begin{equation}
-\ln \mathcal{Z} = 2(\ln \omega_0 + \ln \beta) + \sum_{n=1}^{\infty}\ln X_n,
\end{equation}
where $X_n = \frac{(\nu_n^2+\omega_0^2+\gamma_0\nu_n)^2+(\omega_c\nu_n)^2}{\nu_n^4}$. In the above equation, \(\mathcal{Z}\) is the canonical partition function obtained in the Gibbs approach by evaluating Euclidean (imaginary time) path integrals (see also \cite{PRE79,malay1}). The internal energy is obtained as
\begin{eqnarray}
E=-\frac{\partial \ln \mathcal{Z}}{\partial \beta} =\frac{2}{\beta}+ \sum_{n=1}^{\infty}\frac{1}{X_n}\frac{\partial X_n}{\partial \beta},
\end{eqnarray}
where
\begin{eqnarray}
&&\frac{\partial X_n}{\partial \beta}=-\frac{\nu_n}{\beta}\frac{\partial X_n}{\partial \nu_n}\nonumber \\
&&=\frac{2}{\beta\nu_n^4}\Big[\Big(\nu_n^2+\omega_0^2+\gamma_0\nu_n\Big)(2\omega_0^2+\gamma_0\nu_n)+(\omega_c\nu_n)^2\Big].  \label{intenergygibbs}
\end{eqnarray}
It then follows that Eq. (\ref{intenergygibbs}) matches exactly with Eq. (\ref{int1}) establishing the equivalence between the results obtained via the Einstein approach and the Gibbs approach. One can also obtain the $\gamma_0 =0$ limit from Eq. (\ref{int1}):
\begin{eqnarray}
E_{\gamma_0=0}= \frac{2}{\beta}\Big[1+\sum_{n=1}^{\infty}\frac{(\nu_n^2+\omega_0^2)\omega_0^2+(\nu_n\omega_c)^2}{(\nu_n^2+\omega_0^2)^2+(\nu_n\omega_c)^2}\Big],
\end{eqnarray} 
which matches with Eq. (41) of \cite{sdg2} obtained from an independent calculation of $E$ from the partition function.

\section{Conclusions and discussion}\label{conclusions}
Considering a paradigmatic model of dissipative diamagnetism, we shed light on certain aspects of diamagnetism in open quantum systems. Starting from the quantum
Langevin equation for a dissipative charged particle in a magnetic field, we formulate the quantum counterpart of energy equipartition theorem for the model system in terms of the relaxation function $\Phi(\omega)$ and the universal power spectrum of quantum noise: $u(\omega)=\frac{\hbar\omega}{2}\coth\big(\frac{\hbar\omega}{2k_BT}\big)$ \cite{ford}. The mean kinetic energy of the dissipative system can be expressed in accordance with the quantum equipartition theorem, as integrals involving $[\Phi(\omega)+\Phi(-\omega)]$ and $u(\omega)$. The latter corresponds to the thermally-averaged kinetic/potential energy of a two-dimensional bath oscillator. Following this, we computed the equilibrium magnetic moment of the model system as an integral involving $[\Phi(\omega)-\Phi(-\omega)]$ and the thermal Bohr magneton: $ m(\omega) = -\mu_B \coth\big(\frac{\hbar\omega}{2k_BT}\big)$. Upon putting \(\omega \rightarrow - \omega\) in the integral involving \(\Phi ( - \omega)\), we find that the magnetic moment can be expressed as \(M_z = \langle m(\omega) \rangle\) where \(\langle \cdot \rangle\) implies an average over the probability distribution function \(\mathcal{P}_m(\omega)\) which appears to be related to the quantum counterpart of energy equipartition theorem for the kinetic energy.  \\

We also discuss our results from a superstatistics viewpoint and reformulate the expressions in the energy/magnetic moment representation. Here, the two-fold averaging procedure is as follows: (i) averaging over the thermal Gibbs state for the heat-bath oscillators and, (ii) averaging over energies (or thermal Bohr magneton \(m(\omega)\)). The latter averaging is performed over suitable distribution functions \(f_k(\mathcal{E}_k,T)\) and \(f_m(m,T)\), respectively, for the kinetic energy and magnetic moment of the dissipative oscillator. These distribution functions are sensitive to various parameters such as strength of magnetic field, nature of dissipation mechanism, and strength of coupling between the system and the heat bath. In addition, they also depend upon the temperature of the heat bath. The present investigation on superstatistics appears to be helpful towards understanding the quantum energy partition and it enables us to reinterpret some features of open quantum systems. It should be mentioned that in the same way that we have analyzed the kinetic energy of the dissipative oscillator in this paper, one could analyze the potential energy \cite{kaur}, although we do not pursue it explicitly. \\

Finally, we demonstrate the equivalence between results obtained via the usual Gibbs thermodynamics method and the Einstein approach, based on the quantum Langevin equation. The model system considered here is rather well studied and mimics the realistic three-dimensional case \cite{malay,malay1,20,kaur}. Our results on the (dissipative) orbital magnetic moment can be tested via cold atom experiments with hybrid traps for ions, such as a single ion dipped in a Bose-Einstein condensate \cite{25}. Further, one can generate a uniform magnetic field by utilizing magnetic coils of Helmholtz configuration. The dissipative environment can be built up via 3D optical molasses \cite{26} in combination with a magnetic or an optical trap. One can change the temperature by varying the depth of the
trap and measure the orbital magnetic moment at low temperatures as well as at high temperatures. We hope our work shall stimulate further investigations in this direction.

\subsection*{Acknowledgements}
J.K. gratefully acknowledges financial support from IIT Bhubaneswar in the form of an Institute Research Fellowship. The work of A.G. is supported by Ministry of Education (MoE), Government of India in the form of a Prime Minister's Research Fellowship (ID: 1200454). M.B. is supported by the Department of Science and Technology (DST), Government of India under the Core grant (Project No. CRG/2020/001768) and MATRICS grant (Project no. MTR/2021/000566).

\end{document}